\documentclass[final,1p,times]{elsarticle} 
\usepackage{graphicx} 
\usepackage{amssymb} 
\usepackage{amsthm} 
\usepackage{lineno} 

\journal{Nuclear Physics A} 
\begin{document} 

\begin{frontmatter} 


\title{Effects of Bulk Viscosity on $p_T$-Spectra and Elliptic Flow Parameter}

\author{Akihiko Monnai and Tetsufumi Hirano}

\address{Department of Physics, The University of Tokyo, Tokyo 113-0033, Japan}

\begin{abstract} 
We estimate the effects of viscosity on the phase space distribution appearing in the Cooper-Frye formula within the framework of the Grad's fourteen moment method and find that there are non-trivialities in the discussion of a multi-component system. 
We calculate the viscous corrections of particle spectra and elliptic flow coefficients from the distortion of the distribution using the flow and the hypersurface taken from a (3+1)-dimensional ideal hydrodynamic simulation. 
We see that the bulk viscosity have visible effects on particle spectra. 
The results suggest we should treat \textit{both} shear and bulk viscosity carefully when constraining the transport coefficients from experimental data.
\end{abstract} 

\end{frontmatter} 



\section{Introduction}
Ideal hydrodynamic models have been successful \cite{Hiranov2eta, HT02} for the quark-gluon plasma (QGP) created at the Relativistic Heavy Ion Collider (RHIC) in Brookhaven National Laboratory (BNL). Our next step is to develop viscous hydrodynamic models because the success suggests the deviation from the equilibrium is not so large. We mainly consider the effects of bulk viscosity since the importance of bulk viscosity in terms of the violation of scale invariance near the crossover temperature $T_c$ has been re-recognized \cite{Kharzeev:2007wb}. Particle spectra are affected by viscosity in two ways: one is the variation of the flow $\delta u^\mu = u^\mu - u^\mu _0$ and the hypersurface $\delta d \sigma _\mu = d \sigma _\mu - d \sigma _{0 \mu}$, the other is the distortion of the distribution $\delta f = f -f_0$ because one needs to employ the Cooper-Frye formula 
\begin{equation}
\label{eq:ptspectra}
\frac{d^2N_i}{d^2p_Tdy} = \frac{g_i}{(2\pi )^3} \int _\Sigma p^\mu _i d \sigma _\mu (f^i_0 + \delta f^i)
\end{equation}
at freezeout to convert the hydrodynamic results into the particle picture. Here the index $i$ denotes particle species and $g_i$ the degeneracy. This enables us to follow further dynamical evolution of the hot QCD matter in the cascade models.
In this study we concentrate on corrections to particle spectra due to the modification of the distribution at freezeout in the framework of Grad's method.

\section{Relativistic Kinetic Theory}
We express the distortion of the distribution $\delta f^i$ in terms of macroscopic variables for a relativistic multi-component gas within the framework of the Grad's 14-moment method. For this we generalize the method developed in Ref.~\cite{Israel:1979wp} for a single component system by specifying non-trivialities proper to a system of multi-species particles. The details are shown in Ref. \cite{Monnai:2009ad}.

We have 14 equations that relate the macroscopic variables with the microscopic distribution $\delta f_i$: the kinetic definitions of the bulk pressure $\Pi = -\frac{1}{3} \Delta_{\mu \nu} \delta T^{\mu \nu}$, the energy current $W^\mu = \Delta ^\mu _{\ \alpha} \delta T^{\alpha \beta} u_\beta$, the charge current $V^\mu = \Delta ^\mu _{\ \nu} \delta N^\nu_B$ and the shear stress tensor $\pi ^{\mu \nu} = \delta T ^{\langle \mu \nu \rangle} = [ \frac{1}{2} (\Delta ^\mu _{\ \alpha} \Delta ^\nu _{\ \beta} + \Delta ^\mu _{\ \beta} \Delta ^\nu _{\ \alpha}) - \frac{1}{3}\Delta^{\mu \nu} \Delta_{\alpha \beta}] \delta T^{\alpha \beta}$ and the two Landau matching conditions $u_\mu \delta T^{\mu \nu} u_\nu=0$ and $u_\mu \delta N^{\mu}_B = 0$.
Here $\delta T^{\mu \nu}$ and $\delta N^\mu_B$ are the dissipative parts of the energy-momentum tensor and the baryon number flux. $u^\mu$ is the flow and $\Delta^{\mu \nu} = g^{\mu \nu}-u^\mu u^\nu$ is the projection operator where the metric $g^{\mu \nu} = \mathrm{diag}(+,-,-,-)$. We need the Landau matching conditions to ensure the thermodynamic stability to the first order. This can be shown by explicitly calculating 
$\partial (s^\mu u_\mu)/ \partial \Pi |_{\Pi =0}$, where $s^\mu$ is 
the entropy current. These equations can be written in terms of $\delta f^i$ by using the expressions of $\delta T^{\mu \nu}$ and $\delta N^\mu_B$ in the relativistic kinetic theory for a multi-component system,
\begin{eqnarray}
\label{eq:energy-momentum}
\delta T ^{\mu \nu} = \sum _i \int \frac{g_i d^3 p}{(2\pi )^3 E_i} p_i^\mu p_i^\nu \delta f ^i, \
\label{eq:number}
\delta N ^{\mu }_{B} = \sum _i \int \frac{b_i g_i d^3 p}{(2\pi )^3 E_i} p_i^\mu \delta f ^i ,
\end{eqnarray}
where $b_i$ is the baryon number.

The Grad's 14-moment method allows us to write the distortion of the distribution for the $i$-th particle species with 14 unknowns as
\begin{equation}
\label{eq:df}
\delta f^i = - f_0^i (1 \pm f_0^i) (p_i^\mu \varepsilon _{\mu} + p_i^\mu p_i^\nu \varepsilon _{\mu \nu} ),
\end{equation}
where the positive sign is for bosons and the negative one for fermions.
In the conventional formalism one has the scalar term $\varepsilon$, but here we do not have it because the traceless condition is not imposed on the tensor term $\varepsilon _{\mu \nu}$. These two formalisms are not equivalent for a multi-component system because the scalar term in the conventional method has no particle species dependence whereas the trace part in our method does in the sense that it is dependent on the mass of the $i$-th particle species.

We obtain three independent sets of equations by inserting $\delta f^i$ into the conditions:
\begin{eqnarray}
\left(
 \begin{array}{ccc}
  -J_{30} & -(J_{40}-J_{41}) & -J_{41} \\
  -\tilde{J}_{20} & -(\tilde{J}_{30}-\tilde{J}_{31}) & -\tilde{J}_{31} \\
  J_{31} & (J_{41}-\frac{5}{3}J_{42}) & \frac{5}{3}J_{42}
 \end{array}
\right)
\left(
 \begin{array}{c}
  u^\mu \varepsilon _\mu \\
  u^\mu u^\nu \varepsilon_{\mu \nu} \\
  \mathrm{Tr}(\varepsilon_{\mu \nu})
 \end{array}
\right)
&=&
\left(
 \begin{array}{c}
  0 \\
  0 \\
  \Pi
 \end{array}
\right) ,
\\
\left(
 \begin{array}{cc}
  -J_{31} & - 2J_{41} \\
  -\tilde{J}_{21} & -2\tilde{J}_{31} \\
 \end{array}
\right)
\left(
 \begin{array}{c}
  \Delta^{\mu \nu} \varepsilon_{\nu} \\
  \Delta^{\mu \nu} u^\rho \varepsilon_{\nu \rho} \\
 \end{array}
\right)
 &=&
\left(
 \begin{array}{c}
  W^\mu \\
  V^\mu
 \end{array}
\right) ,
\\
 -2J_{42} \varepsilon ^{\langle \mu \nu \rangle} 
 &=&
 \pi^{\mu \nu} ,
\end{eqnarray}
where $J_{mn}$ is a coefficient of the expansion of the moment
\begin{eqnarray}
J^{\mu _1 ... \mu _m} & = & \sum _i \int \frac{g_i d^3 p}{(2\pi )^3 E_i} f_0^i (1 \pm f_0^i) p_i^{\mu _1} ... p_i^{\mu _m}\\
& = & \sum _n J_{mn} \big[ (\Delta ^{\mu _1 \mu _2} ... \Delta ^{\mu _{2n-1} \mu _{2n}} u^{\mu _{2n+1}} ... u^{\mu _m}) + \mathrm{(permutations)} \big].
\end{eqnarray}
$\tilde{J}_{mn}$ is defined in the same way but with the baryon number $b_i$ as an additional weight factor in the summation over the particle species index $i$. The unique form of $\delta f^i$ can be obtained by solving them. The correction tensors are
\begin{eqnarray}
\label{eq:linear}
\varepsilon _{\mu} &=& D_0 \Pi u_\mu + D_1 W_\mu + \tilde{D}_1 V_\mu , \\
\varepsilon _{\mu \nu} &=& (B_0 \Delta _{\mu \nu} + \tilde{B}_0 u_\mu u_\nu )\Pi + 2 B_1 u_{( \mu} W_{\nu )} + 2 \tilde{B}_1 u_{( \mu} V_{\nu )} + B_2\pi_{\mu \nu},
\end{eqnarray}
where the prefactors $D_i$ and $B_i$ are functions of $J_{mn}$ and $\tilde{J}_{mn}$. One cannot take the scalar term in (\ref{eq:df}) because that induces a change of the sign in the bulk pressure-related prefactors $D_0$, $B_0$ and $\tilde{B}_0$ which causes divergence in the case of the 16-component hadron resonance gas discussed later. Note here that (i) information of all the components in the gas comes into the viscous modification of a particle species through the transport coefficients, $J_{mn}$'s and $\tilde{J}_{mn}$'s, and (ii) if one takes the quadratic ansatz in Ref. \cite{Dusling}, \textit{i.e.}, $\varepsilon _{\mu \nu} = C_1 \pi _{\mu \nu} +C_2 \Delta _{\mu \nu} \Pi$, the system becomes thermodynamically unstable because the matching conditions are not satisfied.

\section{Results}
\begin{figure}[ht]
\centering
\includegraphics[width=0.49\textwidth]{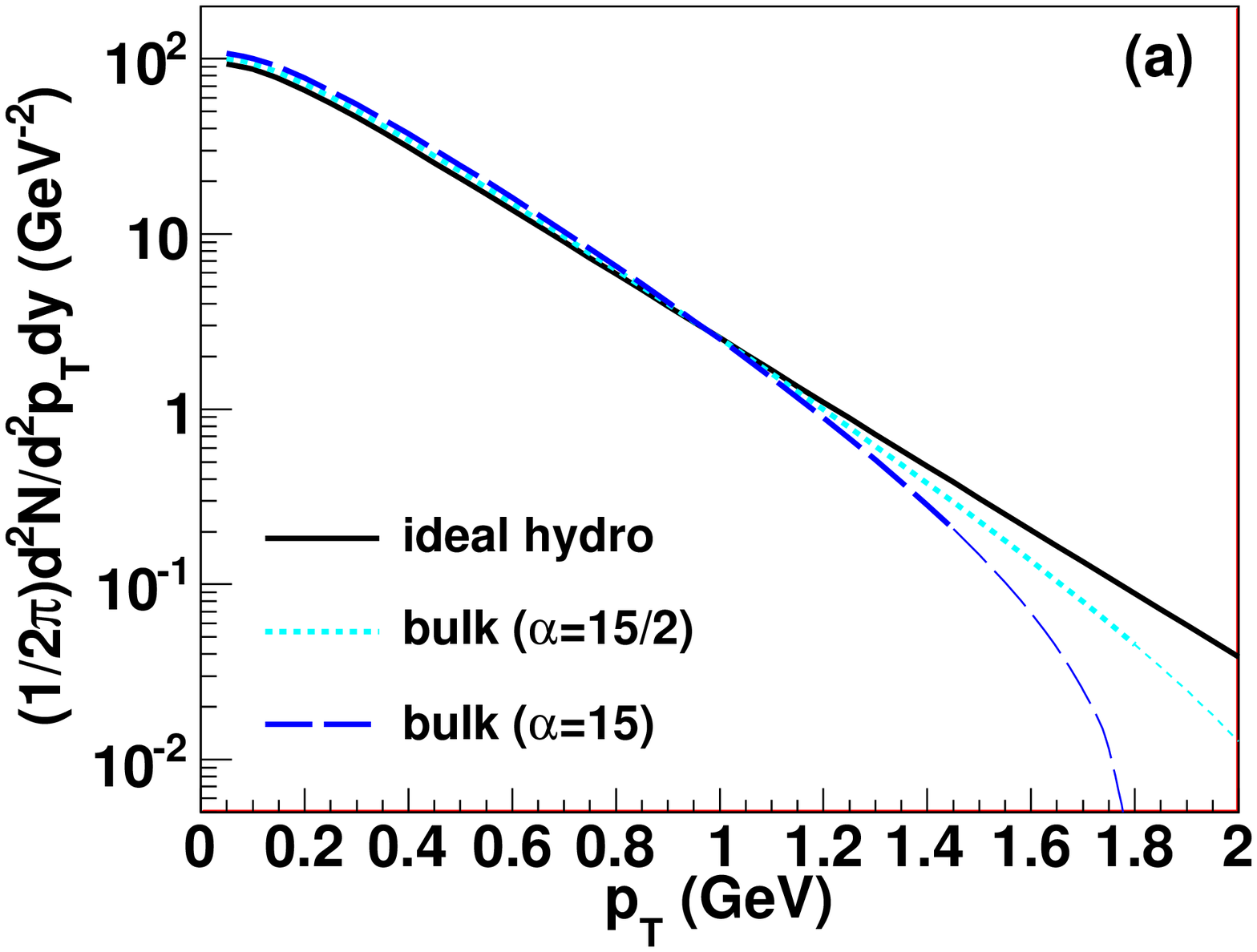}
\includegraphics[width=0.49\textwidth]{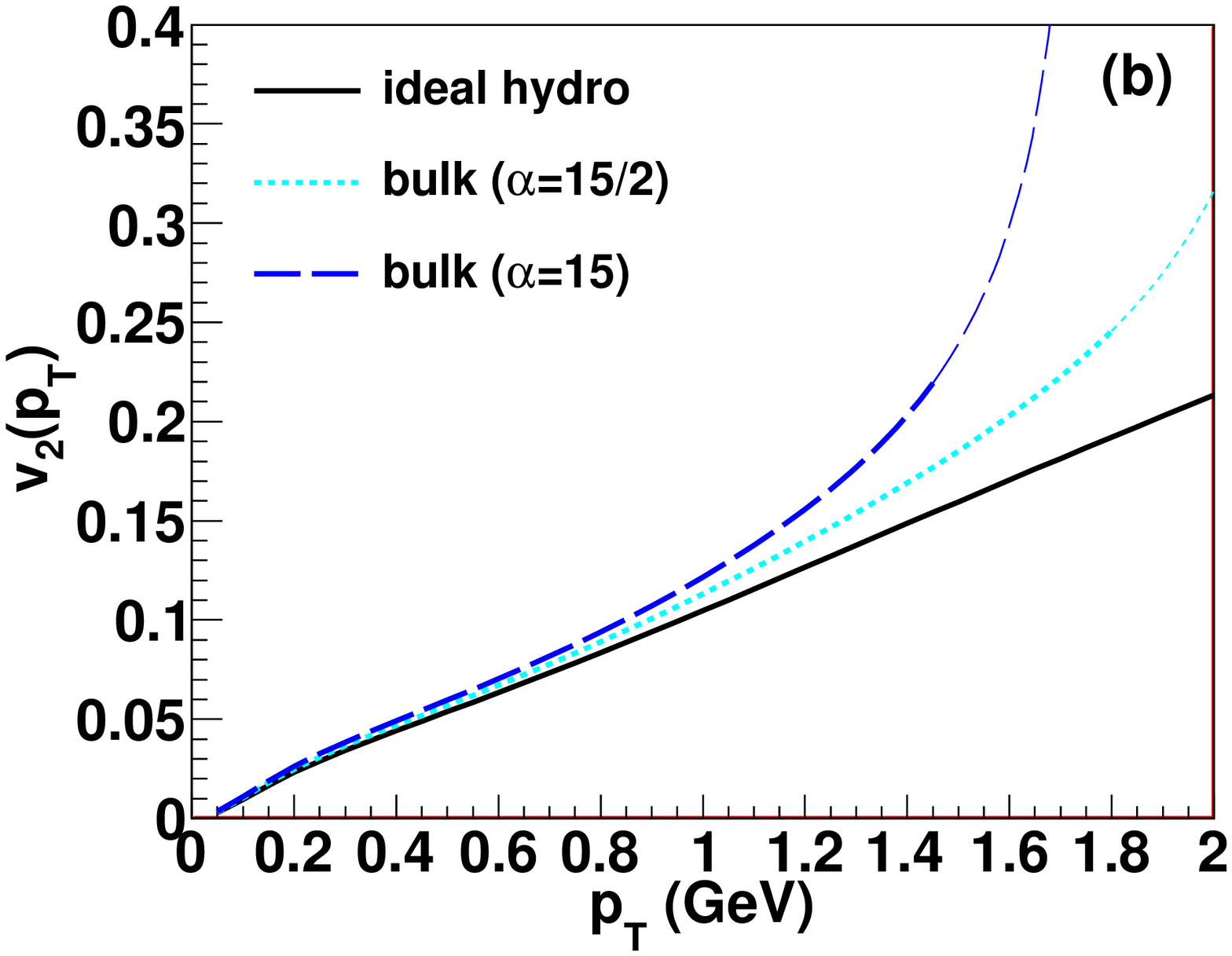}
\caption[]{Effects of bulk viscosity on (a) $p_T$-spectra and (b) elliptic flow coefficients $v_2(p_T)$ for the negative pion. Solid, dotted, and dashed lines are, respectively, the results without any viscous corrections, with the effect of bulk viscosity with $\alpha = 15/2$ and with $\alpha = 15$. The thick(thin) lines show that the absolute value of the ratio of the correction to the ideal spectrum is smaller(larger) than 0.5.
}
\label{fig:result}
\end{figure}
One needs models for the equation of state (EoS), the macroscopic variables and profiles of the flow and the hypersurface to estimate the effects of $\delta f$ on particle spectra.
We consider 16-component hadron resonance gas for the EoS, which has mesons and baryons with the mass up to $\Delta (1232)$. 
The bulk pressure is estimated with the Navier-Stokes limit, \textit{i.e.}, $\Pi = -\zeta \partial _\mu u^\mu$ where the models of bulk and shear viscosity are $\zeta = \alpha (c_s^2 -\frac{1}{3})^2 \eta$ \cite{Weinberg:1971mx, Hosoya:1983id, Arnold:2006fz} and $\eta = \frac{s}{4\pi}$ \cite{Son_visc}. Here $s$ and $c_s$ are the entropy density and the sound velocity, respectively. $\alpha$ is a free parameter. The freezeout temperature $T_f$ is set to $0.160$ GeV where $\eta = 1.31\times 10^{-3}\ \mathrm{GeV}^3$ and $\zeta = 4.37\times 10^{-4}\ \mathrm{GeV}^3$ when $\alpha = 15$. 
The flow and the hypersurface are taken from a (3+1)-dimensional ideal hydrodynamic simulation \cite{Hiranov2eta, HT02} in Au+Au collisions at $\sqrt{s_{NN}} = 200$ GeV where the impact parameter is 7.2 fm. It should be noted again that what we estimate here is not the distortion of the flow $\delta u^\mu$ but the viscous corrections through the modification of the distribution $\delta f$ at freezeout.
For numerical estimations we employ the zero net baryon density limit, which is well justified for the hot QCD matter at RHIC.
We do not have to worry about the loss of the Landau matching condition of the baryon number flux because the condition still yields a finite relation even in the limit of vanishing chemical potential \cite{Monnai:2009ad}.

We estimate the particle spectra of negative pion with mass $m = 0.139$ GeV using the Cooper-Frye formula (\ref{eq:ptspectra}) without changes of flow velocity nor freezeout hypersurface from the ideal hydrodynamic results. The bulk viscosity lowers the mean $p_T$ of the particle spectra, $\langle p_T \rangle$, when estimated with the ideal hydrodynamic flow and hypersurface. This can be attributed to the reduction of the effective pressure due to the bulk viscosity. The isotropic nature of the bulk pressure reduces the pressure in transverse directions when the system is expanding in longitudinal direction. On the other hand, the bulk viscous correction enhances the elliptic flow coefficients $v_2(p_T)$. This can be interpreted in the relation $dv_2(p_T)/dp_T \simeq v_2/\langle p_T \rangle$ \cite{HG05}, because the slope of differential $v_2(p_T)$ is expected to become steep when the $\langle p_T \rangle$ is decreased while the average $v_2$ is not affected by viscosity with
 in our calculations.

We have to be aware that the viscous corrections might have been overestimated because (i) we take the Navier-Stokes limit for $\Pi$ and thus do not consider relaxation effects, and (ii) the derivatives of ideal hydrodynamic flow is generally larger than those of real viscous hydrodynamic flow as viscosity tends to eliminate differences in the flow.

\section{Conclusions}

We obtained the unique and consistent expression of the distortion of the distribution for a multi-component system in the framework of the Grad's 14-moment method. We find several non-trivial issues along the course of discussion. First, one needs to introduce the non-zero trace tensor term in the correction instead of the conventional scalar term in the case of the 16-component hadron resonance gas to get a finite result. Second, the distortion is different by particle species, but is dependent on all the components in the gas because the transport coefficients and macroscopic quantities $J_{mn}$ and $\tilde{J}_{mn}$ are. Third, the matching for the baryon number flux remains meaningful even in the limit of vanishing chemical potential.

Effects of bulk viscosity on particle spectra through the distortion of the distribution is estimated, and we find it suppresses $p_T$-spectra and enhances $v_2(p_T)$ when estimated with (3+1)-dimensional ideal hydrodynamic flow and hypersurface. We should be careful of the possible overestimation in the usage of the Navier-Stokes limit and the ideal hydrodynamic flow. A (3+1)-dimensional viscous hydrodynamic model is necessary to accurately estimate the viscous effects to compare with experimental data and to constrain the transport coefficients from them.


\section*{Acknowledgments}
The authors acknowledge fruitful discussions with T.~Hatsuda, T.~Kodama, T.~Kunihiro, and S.~Muroya.
The authors are also grateful for valuable comments by G.~S.~Denicol.
The work of T.H. was partly supported by Grant-in-Aid for Scientific Research No.~19740130 and by Sumitomo Foundation No.~080734.

\end{document}